\documentclass[12pt]{article}
\usepackage{a4}

\usepackage{amsmath}
\usepackage{amsfonts}
\usepackage{fancybox}

\usepackage{cite}
\usepackage{bm}
\allowdisplaybreaks

\newcounter{A}

\usepackage[dvips]{color}

\def\Tr{\mathrm{Tr}}

\def\diag{\mathrm{diag}}

\def\[{\left[}
\def\]{\right]}
\def\({\left(}
\def\){\right)}

\def\cN{{\cal N}}
\def\cO{{\cal O}}

\def\bS{{\mathbf S}}
\def\bR{{\mathbf R}}
\def\bZ{{\mathbf Z}}

\def \be {\begin{equation}}
\def \ee {\end{equation}}
\def \bea {\begin{eqnarray}}
\def \eea {\end{eqnarray}}

\newcommand{\n}{\nonumber \\}

\begin{document}




\vskip 1.5cm

\setlength{\oddsidemargin}{0cm}
\setlength{\baselineskip}{7mm}

\begin{titlepage}
\renewcommand{\thefootnote}{\fnsymbol{footnote}}
\begin{normalsize}
\begin{flushright}
\begin{tabular}{l}
\\
SNUTP11-006\\
HRI/IST/1105 \\
KUNS-2349
\end{tabular}
\end{flushright}
  \end{normalsize}

~~\\

\vspace*{0cm}
    \begin{Large}
    \begin{bf}
       \begin{center}
         {BPS solutions in ABJM theory and \\ Maximal Super Yang-Mills on $\bR\times \bS^2$}
       \end{center}
    \end{bf}   
    \end{Large}
\vspace{0.7cm}

\begin{center}
Bobby Ezhuthachan$^{1)}$\footnote
            {
e-mail address : 
bobby(at)hri.res.in}, 
Shinji Shimasaki$^{1),2)}$\footnote
            {
e-mail address : 
shinji(at)gauge.scphys.kyoto-u.ac.jp}
    and
Shuichi Yokoyama$^{3),4)}$\footnote
           {
e-mail address : 
yokoyama(at)phya.snu.ac.kr}\\

\vspace{0.7cm}
                    
 $^{1)}$ {\it Harish-Chandra Research Institute,
Chhatnag Rd, Jhunsi, Allahabad 211019, India}\\
 \vspace{0.3cm}
$^{2)}$ {\it Department of Physics, Kyoto University, Kyoto 606-8592, Japan}\\
\vspace{0.3cm}
 $^{3)}$ {\it Department of Physics and Astronomy and Center for Theoretical Physics, \\
Seoul National University, Seoul 51-747, Korea}\\
\vspace{0.3cm}
 $^{4)}$ {\it Department of Physics, University of Tokyo,
Tokyo 113-0033, Japan}\\  
               
\end{center}

\vspace{0.7cm}

\begin{abstract}
\noindent

We investigate BPS solutions in ABJM theory on $\bR\times \bS^2$.
We find new BPS solutions, 
which have nonzero angular momentum as well as nontrivial configurations of fluxes. 
Applying the ``Higgsing procedure'' of arxiv:0803.3218 around a $1/2$-BPS solution
of ABJM theory, one obtains $\cN=8$ super Yang-Mills (SYM) on $\bR\times \bS^2$. 
We also show that other BPS solutions of the SYM can be obtained
from BPS solutions of ABJM theory by this higgsing procedure.


\end{abstract}
\vfill
\end{titlepage}
\vfil\eject

\setcounter{footnote}{0}


\section{Introduction}
\setcounter{equation}{0}

Superconformal Chern-Simons-matter (CSM) theories have been studied with 
considerable interest over the past few years. These theories have been 
studied in the context of M-theory and their possible relevance to the world-volume 
theory of multiple M2-branes was first discussed in \cite{Schwarz:2004yj}. 
The first explicit Lagrangian of such a CSM theory was BLG theory
\cite{Bagger:2006sk, Bagger:2007jr, Bagger:2007vi, Gustavsson:2007vu}. 
This was a maximally supersymmetric $\mathcal{N}=8$ superconformal theory of fixed rank 
$SU(2)\times SU(2)$ coupled to matter fields transforming in the bi-fundamental of 
the two $SU(2)$'s. 
The Chern-Simons terms of the two $SU(2)$'s come with a relative negative sign. 
Even though the relevance of the BLG theory to M2-brane theory is not understood, 
CSM theories with lesser supersymmetry, sharing some of 
the above mentioned features of the BLG theory, 
have been proposed as the world-volume description of M2-branes in various backgrounds. 
In particular, a certain $\mathcal{N}=6$ superconformal CSM theory - ABJM theory - 
was proposed as the world-volume theory of multiple M2-branes on $\mathbb{C}^4/\mathbb{Z}_{k}$,
where $k$ is the Chern-Simons level \cite{Aharony:2008ug}. 
For $k=1,2$, ABJM theory has $\mathcal{N}=8$ supersymmetries
 even though in the classical Lagrangian only $\mathcal{N}=6$ supersymmetries are manifest. 
The enhanced symmetry generators are realized in terms of monopole operators
\cite{Aharony:2008ug,Gustavsson:2009pm,Bashkirov:2010kz}.

Several checks have been done for this proposal. Firstly the  moduli space 
of the theory has been shown to have the right geometry. 
In the case of ABJM theory, for instance, 
the moduli space is $\mathbb{C}^4/\mathbb{Z}_{k}$. 
Tests beyond getting the right moduli space have also been done. 
This includes the computation of the superconformal index of the theory 
and matching with results from supergravity
\cite{Bhattacharya:2008bja, Imamura:2008ji,Imamura:2009ur, Kim:2009wb, Imamura:2009hc}. 
Several CSM theories have been proposed to describe M2-branes in other backgrounds
\cite{Hosomichi:2008jb, Aharony:2008gk, Jafferis:2008qz, Martelli:2008si, 
Martelli:2008rt, Hanany:2008cd, Hanany:2008fj,Franco:2008um, Franco:2009sp}. 

One of the first checks of the relevance of these CSM theories to M-theory 
was performed in \cite{Mukhi:2008ux, Distler:2008mk}. 
In the case of M2-branes on $\mathbb{C}^4/\mathbb{Z}_{k}$, 
one can consider a limit in which we take the branes far away from the orbifold fixed point 
and simultaneously take small orbifold angle. 
In this limit the orbifold geometry can be well approximated by $\bS^1 \times \bR^7$. 
This is the limit 
in which the M2-branes should be approximated by D2-branes, and therefore 
the CSM theory should be approximated by a super Yang-Mills theory (SYM). 
Mukhi and Papageorgakis gave a field theory realization of this picture 
in BLG theory\footnote{
Even though the geometry of the moduli space of BLG theory is more complicated than 
$\mathbb{C}^{4}/\mathbb{Z}_{k}$, the Higgsing procedure still leads to SYM.}. 
By giving a vev to a scalar field, and taking the large $v$ and large $k$ limit 
with $\frac{2\pi v^2}{k}=g^{2}_{ym}$ held constant as the gauge coupling, 
it was shown that the CSM theory is approximated by $\mathcal{N}=8$ SYM 
on flat spacetime. 
This procedure was called the ``novel Higgs mechanism''. 
This was first done in the context of the maximally supersymmetric 
$\mathcal{N}=8$ BLG theory but carries 
over for ABJM theory as well\cite{Aharony:2008ug}.

For the abelian versions of the theories, 
corresponding to a single D2 brane and single M2 brane, 
it can be explicitly seen that the ABJM at $k=1$ can be rewritten as the SYM 
by simply compactifying one of the eight-scalar fields and dualizing it into a gauge field. 
Of course, for the non-abelian theory, it is not possible to carry out a compactification 
directly at the level of the classical Lagrangian 
because the translation invariance along the transverse directions is not manifest 
in the Lagrangian. 
Also, since the SYM is interacting, one expects the $SO(8)$ invariance to be manifest
only at the strongly coupled IR fixed point of the SYM\footnote{
However, in \cite{Agarwal:2011tz}, 
it was shown that even in the non-abelian case the enhanced $SO(8)$ invariance 
can be seen manifestly at the level of scattering amplitudes of the SYM. 
See also \cite{Agarwal:2011hb}.}. 
Therefore the Higgsing procedure is the only way in which one can see the M2 to D2 connection 
at the level of the classical Lagrangian.

Since ABJM theory is conformal there exists a conformal map which 
maps ABJM theory on flat spacetime to that on $\bR\times \bS^2$. Under this 
map the vacua of ABJM theory get mapped to time-dependent 
$1/2$-BPS solutions on $\bR\times \bS^2$ \cite{Berenstein:2009sa}. 
The novel Higgs mechanism was carried out around the vacua of the CSM theory on 
flat space and resulted in $\mathcal{N}=8$ SYM. 
It is worth asking what happens when we carry out the analogous procedure of 
the novel Higgs mechanism about the corresponding solutions of ABJM theory 
on $\bR\times\bS^2$. 
In this case, it is naturally expected that we obtain $\cN=8$ SYM\footnote{
$\mathcal{N}=8$ SYM on $\bR\times \bS^2$ 
is no longer related to the $\mathcal{N}=8$ SYM on flat space 
because the theory is not conformal.
} on $\bR\times \bS^2$,
which preserves $SU(2|4)$ symmetry (16 supersymmetries) and 
has been studied previously 
in the context of the plane wave (BMN) matrix model \cite{Berenstein:2002jq},
gauge/gravity duality \cite{Lin:2005nh,Ishiki:2006yr}
and the large-$N$ reduction of $\cN=4$ SYM on $\bR\times \bS^3$ \cite{Ishiki:2006yr}.
Thermodynamic aspects of this SYM was studied in \cite{Grignani:2007xz}
while aspects related to integrability was studied in \cite{Agarwal:2010tx}.

In this paper, we first solve for BPS configurations in ABJM theory on $\bR\times \bS^2$.
In particular, we find general BPS solutions for diagonal configurations.
Interestingly, 
the BPS solutions have non-trivial $(t,\theta, \varphi)$-dependence on $\bR\times \bS^2$ 
with nonzero angular momentum on $\bS^2$
as well as non-trivial flux, not only ``magnetic flux'' but also ``electric flux'', turned on. 
We then show that carrying out the Higgsing procedure 
around a $1/2$-BPS solution of ABJM theory on $\bR\times \bS^2$ 
leads to $\mathcal{N}=8$ SYM on $\bR\times \bS^2$.
In this process, as in the flat space case,
we observe an enhancement of the supersymmetry and the $R$-symmetry, 
from 12 and $SU(3)$ to 16 and $SU(4)$, respectively\footnote{
This is the supersymmetry and global symmetry preserved by the 
1/2-BPS solution about which we ``Higgs''.}.
We also comment on the mechanism of this enhancement.
Furthermore we show that the theory around a nontrivial vacuum and a $1/2$-BPS solution of 
$\mathcal{N}=8$ SYM on $\bR\times \bS^2$ is also obtained by Higgsing
the theory around another $1/2$-BPS solution and a $1/4$-BPS solution, respectively, 
of ABJM theory on $\bR\times \bS^2$.

%

The organization of this paper is as follows.
In section 2, we write down the action, equations of motion and supersymmetries of 
 ABJM theory on $\bR\times \bS^2$.
In section 3, we solve for specific $1/2$-BPS and $1/4$-BPS solutions of this theory. 
In section 4, we then show that higgsing around 
a $1/2$-BPS solution of ABJM on $\bR\times \bS^2$ leads to 
the $\mathcal{N}=8$ SYM on $\bR\times \bS^2$ and make some comment on the symmetry enhancement. 
We also show that theories expanded around a nontrivial vacuum and a $1/2$-BPS solution 
of $\mathcal{N}=8$ SYM on $\bR\times \bS^2$ are obtained from ABJM theory. 
Section 5 is devoted to summary and discussion.
There are four appendices in which we collect our notations 
and conventions used in the paper, give some details about the BPS solutions 
of ABJM theory on $\bR\times \bS^2$, 
present the action, supersymmetry transformations and vacuum solutions 
of the $\mathcal{N}=8$ SYM on $\bR\times \bS^2$
and give some details about the representation of the $R$-symmetry 
of fermions in ABJM theory and SYM.   
 

\section{ABJM on $\bf R\times S^2$}
\setcounter{equation}{0}

In this section we write down the action, equations of motion 
and supersymmetry transformations of ABJM theory on $\bf R\times S^2$ 
with Minkowski signature $(-++)$. 

The field content of ABJM theory is the following: two gauge fields 
$A^{(1)}$ and $A^{(2)}$ associated with the gauge group $U(N)\times U(N)$, 
bi-fundamental scalars $Y^A$ and their superpartners $\psi_A$ ($A=1,2,3,4$),
which are $(1+2)$-dimensional Majorana spinors. 
The global symmetry of this theory is the superconformal symmetry $OSp(6|4)$
and a $U(1)$ (baryon) symmetry, denoted by $U(1)_b$.  
$OSp(6|4)$ includes the $(1+2)$-dimensional conformal group $SO(2,3)$ 
and $R$-symmetry $SU(4)$ as bosonic subgroups.
 $Y^A$ ($\psi_A$) transforms as the (anti-)fundamental representation of $SU(4)$
and carries charge -1(+1) under $U(1)_b$.

The action of ABJM theory on $\bf R\times S^2$ is given by
\begin{align}
S&=\int dt \frac{d\Omega_2}{\mu^2} 
\Tr\biggl[
\frac{k}{4\pi}\epsilon^{mnp}\Bigl(
A^{(1)}_m \partial_n A^{(1)}_{p}+\frac{2i}{3}A^{(1)}_{m}A^{(1)}_{n}A^{(1)}_{p}
-A^{(2)}_m \partial_n A^{(2)}_{p}-\frac{2i}{3}A^{(2)}_{m}A^{(2)}_{n}A^{(2)}_{p}
\Bigr) \n
&\qquad
-D_m Y_A^\dagger D^m Y^A -\frac{\mu^2}{4}Y_A^\dagger Y^A
+i\psi^{\dagger A} \gamma^a D_a \psi_A \n
&\quad
+\frac{4\pi^2}{3 k^2} \left(
Y^AY_A^\dagger Y^BY_B^\dagger Y^CY_C^\dagger 
+Y_A^\dagger Y^A Y_B^\dagger Y^B Y_C^\dagger Y^C
+4Y^AY_B^\dagger Y^CY_A^\dagger Y^BY_C^\dagger
-6Y^AY_B^\dagger Y^BY_A^\dagger Y^CY_C^\dagger  
\right) \n
&\quad
+\frac{2\pi i}{k}\left(
\psi_A\psi^{\dagger A}Y^BY_B^\dagger-\psi^{\dagger A}\psi_AY_B^\dagger Y^B
+2\psi^{\dagger A}\psi_BY_A^\dagger Y^B-2\psi_A\psi^{\dagger B}Y^AY_B^\dagger
\right) \n
&\quad
+\frac{2\pi i}{k}\left(
\epsilon_{ABCD}\psi^{\dagger A}Y^B\psi^{\dagger C}Y^D
-\epsilon^{ABCD}\psi_AY_B^\dagger \psi_CY_D^\dagger
\right)\biggr].
\label{ABJM}
\end{align}
where $m,n,p\cdots$ run over the world-volume coordinates $t,\theta,\varphi$ 
and $a,b,\cdots=1,2,3$ are corresponding local Lorentz indices.
The upper and lower $A,B,\cdots$ are indices of $\bm{4}$ and $\bar{\bm{4}}$, respectively, 
of $SU(4)$ and run $1,2,3,4$.
$k(=1,2,\cdots)$ is the Chern-Simons coupling and $\mu^{-1}$ is the radius of $S^2$.
$\gamma^a$ $(a=1,2,3)$ are gamma matrices of $SO(1,2)$, which satisfy 
$\{\gamma^a,\gamma^b\}=2\eta^{ab}$ with $\eta^{ab}=\diag(-1,+1,+1)$.
The mass term of the scalar field comes from the coupling to the background curvature.
Covariant derivatives take the following form
\begin{align}
D_m Y^A&=\partial_m Y^A +iA^{(1)}_{m}Y^A-iY^A A^{(2)}_{m}, \n
D_m \psi_A
&=\nabla_m\psi_A+iA^{(1)}_{m}\psi_A-i\psi_A A^{(2)}_{m} \n
&=\partial_m \psi_A + \frac{1}{4}\omega_{m ab}\gamma^{ab}\psi_A
+iA^{(1)}_{m}\psi_A-i\psi_A A^{(2)}_{m}.
\end{align}
where $\omega_{ab}$ is the spin connection of $\bR\times \bS^2$. 
In appendix A, we gather our conventions of the metric and the spinor used in this paper.
Equations of motion for the bosonic fields with $\psi_A=0$, 
which are relevant for the following discussion, are given by
\begin{align}
\epsilon^{abc}\frac{k}{4\pi}F^{(1)}_{bc}
&=i\left(Y^A D^a Y_A^\dagger -D^a Y^A Y_A^\dagger \right), \n
\epsilon^{abc}\frac{k}{4\pi}F^{(2)}_{bc}
&=i\left(D^a Y_A^\dagger Y^A-Y_A^\dagger D^aY^A\right), \n
\left(D_aD^a-\frac{\mu^2}{4}\right)Y^A
&=-\frac{4\pi^2}{k^2}
\Bigl(
   Y^B Y_B^\dagger Y^C Y_C^\dagger Y^A +  Y^A Y_B^\dagger Y^B Y_C^\dagger Y^C
+4 Y^B Y_C^\dagger Y^A Y_B^\dagger Y^C \n
&\qquad
-2 Y^B Y_B^\dagger Y^A Y_C^\dagger Y^C
-2 Y^A Y_B^\dagger Y^C Y_C^\dagger Y^B -2 Y^B Y_C^\dagger Y^C Y_B^\dagger Y^A 
\Bigr). \label{EOM}
\end{align}

We can show that the action (\ref{ABJM}) is invariant under 
the following supersymmetry transformations\footnote{
For $k=1,2$, there are additional supersymmetries which are not manifest 
in the Lagrangian.} 
\begin{align}
\delta Y^A&= -i \xi^{AB} \psi_B, \n 
\delta Y_A^\dagger
&=-i \psi^{\dagger B}\xi_{AB}, \n
\delta \psi_A
&=-\gamma^m \xi^{}_{AB} D_m Y^B
-\frac{2\pi}{k} Q^{B\phantom{A}C}_{\phantom{B}A} \xi_{BC}
-\frac{1}{3}Y^B\gamma^m \nabla_m \xi_{AB}, \n
\delta \psi^{\dagger A}
&=\xi^{AB}\gamma^m D_m Y_B^\dagger
-\frac{2\pi}{k} (Q^{B\phantom{A}C}_{\phantom{B}A})^\dagger \xi^{BC}
+\frac{1}{3}Y_B^\dagger \nabla_m \xi^{AB}\gamma^m, \n
\delta A^{(1)}_m
&=-\frac{2\pi}{k}\left[
Y^B\psi^{\dagger A} \gamma_m \xi_{AB}
+\xi^{AB}\gamma_m \psi_A Y_B^\dagger \right], \n
\delta A^{(2)}_m
&=-\frac{2\pi}{k}\left[
\psi^{\dagger A}\gamma_m \xi_{AB}Y^B
+Y_B^\dagger \xi^{AB}\gamma_m \psi_A \right], 
\label{SUSY transf}
\end{align}
where 
\begin{align}
&Q^{B\phantom{A}C}_{\phantom{B}A}
\equiv T^{B\phantom{A}C}_{\phantom{B} A}
-\frac{1}{2}\delta_A^C T^{B\phantom{D}D}_{\phantom{B} D}
+\frac{1}{2}\delta_A^B T^{C\phantom{D}D}_{\phantom{C} D},
\quad
T^{B\phantom{A} C}_{\phantom{B} A}\equiv Y^BY_A^\dagger Y^C - Y^C Y_A^\dagger Y^B.
\end{align}
$\xi_{AB}$ are supersymmetry parameters, 
which are $(1+2)$-dimensional Majorana spinors 
and antisymmetric in $A$ and $B$ (i.e. $\bm{6}$ of $SU(4)_R$),
$\xi_{AB}=-\xi_{BA}$, and satisfy the conformal Killing spinor equations,
\begin{align}
\nabla_a\xi_{AB}=\pm i \frac{\mu}{2}\gamma_a\gamma^0 \xi_{AB}.
\label{eq of xi}
\end{align}
Hereafter we denote $\xi_{AB}$ satisfying the upper and lower signs in (\ref{eq of xi})
by $\xi^{(+)}_{AB}$ and $\xi^{(-)}_{AB}$, respectively.
$\xi^{(\pm)AB}$ is the complex conjugate of $\xi^{(\pm)}_{AB}$ and satisfy
\begin{align}
\xi^{(\pm)AB}\equiv (\xi^{(\pm)}_{AB})^*=-\frac{1}{2}\epsilon^{ABCD}\xi^{(\mp)}_{CD}.
\label{cc of xi}
\end{align}
So, $\xi^{(\pm)}_{AB}$ are related to the complex conjugate of $\xi^{(\mp)}_{AB}$.
One can easily solve (\ref{eq of xi}) as
\begin{align}
\xi^{(\pm)}_{AB}
=
e^{\pm i\frac{\mu t}{2}}e^{\mp i\gamma^2 \frac{\theta}{2}} e^{\gamma^0 \frac{\phi}{2}} 
\eta^{(\pm)}_{AB}
 \label{xi in terms of eta}
\end{align}
where $\eta^{(\pm)}_{AB}$ are constant spinors.
Thus the action (\ref{ABJM}) possesses 24 supersymmetries.

\section{BPS solutions of ABJM on $\bR\times \bS^2$}
\setcounter{equation}{0}

In this section, we find specific BPS solutions of ABJM theory on $\bR \times \bS^2$.
BPS solutions, in general, are obtained by solving $\delta \psi_A=0$ 
as well as the equations of motion with $\psi_A=0$.
Since it is difficult to solve the equations generically,
we look for solutions with diagonal configuration in the $U(N)\times U(N)$ theory. 
For these solutions, $Q^{B\phantom{A}C}_{\phantom{B}A}=0$. 
Therefore each diagonal component is basically a BPS solution of the $U(1)\times U(1)$ theory.
The BPS equations can be easily solved with this assumption. 
In the following, we give particular BPS solutions, 
which are $1/2$-BPS and $1/4$-BPS solutions 
for $U(1)\times U(1)$ ABJM theory when $k>2$. 
They are determined by $\delta \psi_A=0$,
where $\delta \psi_A$ is given in \eqref{SUSY transf}.
Other BPS solutions are summarized in appendix B.

\subsection{1/2-BPS solution}

We first look for $1/2$-BPS solutions of ABJM theory on $\bR\times \bS^2$ 
\cite{SheikhJabbari:2009kr,Berenstein:2008dc,Berenstein:2009sa}.
Let us consider the equation given by $\delta \psi_A=0$ in $U(1)\times U(1)$ ABJM theory,
\begin{align}
-\gamma^m \xi^{(\pm)}_{AB} D_m Y^B
\mp i\frac{\mu}{2}Y^B\gamma^0 \xi^{(\pm)}_{AB}=0,
\label{del psi U(1)}
\end{align}
where $\xi^{(\pm)}_{AB}$ is explicitly given in \eqref{xi in terms of eta}.
Since the equations of motion for the gauge fields imply $F^{(1)}_{mn}=F^{(2)}_{mn}$,
we can take a gauge in which 
\begin{align}
A^{(1)}_m=A^{(2)}_m,
\end{align}
so that $D_m$ becomes $\partial_m$ in \eqref{del psi U(1)}.
Now, we look for BPS solutions preserving $SU(3)$ of the $SU(4)$ $R$-symmetry.
Such a configuration is obtained by imposing
\begin{align}
&\eta^{(+)}_{A'B'}=0, \quad \eta^{(+)}_{A'4}\neq 0, \n
&\eta^{(-)}_{A'4}=0, \quad \eta^{(-)}_{A'B'}\neq 0
\label{1/2 BPS condition}
\end{align}
where $A',B',\cdots =1,2,3$ 
and the second line of \eqref{1/2 BPS condition} is the complex conjugate of the first line. 
This is a $1/2$-BPS condition.
Then, \eqref{del psi U(1)} reduces to the equations for the scalars
\begin{align}
&Y^1=Y^2=Y^3=0, \n
&(\partial_t + i\frac{\mu}{2})Y^4=0, \quad 
\partial_\theta Y^4=\partial_\varphi Y^4=0.
\end{align}
Therefore, a $1/2$-BPS solution for the scalar fields is given by
\begin{align}
&Y^1=Y^2=Y^3=0, \n
&Y^4=v e^{-i\frac{\mu}{2}t},
\label{1/2 BPS scalar field}
\end{align}
where $v$ is a complex constant.
This solution breaks $SU(4)$ $R$-symmetry to $SU(3)$.
It turns out from the equations of motion of the gauge fields in \eqref{EOM} that 
the gauge fluxes take the form 
\begin{align}
&F^{(1)}_{01}=F^{(2)}_{01}=F^{(1)}_{02}=F^{(2)}_{02}=0, \n
&F^{(1)}_{12}=F^{(2)}_{12}=\frac{2\pi\mu}{k} |v|^2.
\label{1/2 BPS field strength}
\end{align}
Flux quantization condition;
\begin{align}
 \frac{1}{2\pi}\int \frac{d\Omega}{\mu^2} F^{(i)}_{12}  \in \bm{Z}.
\label{flux quantization}
\end{align}
leads to the quantization of $v$; 
\begin{align}
\frac{4\pi}{\mu k} |v|^2 =2q \in \bm{Z}_{\geq 0},
\label{1/2 BPS quantization condition}
\end{align}
where $q\in \bm{Z}_{\geq 0}/2$.
One can easily solve \eqref{1/2 BPS field strength} locally in terms of gauge fields
by introducing two patches on $S^2$;
\begin{align}
&A^{(1)}_0=A^{(2)}_0=0, \n
&A^{(1)}_1=A^{(2)}_1=0, \n
&A^{(1)}_2=A^{(2)}_2=\frac{2\pi |v|^2}{k}\frac{\pm 1 - \cos\theta}{\sin\theta}
=\mu q \frac{\pm 1 - \cos\theta}{\sin\theta},
\label{1/2 BPS gauge field}
\end{align}
where we have taken $A_0^{(1)}=A_0^{(2)}=A_1^{(1)}=A_1^{(2)}=0$ gauge. 
The upper and lower signs in the third line 
correspond to the region I $(0\leq \theta <\pi)$ 
and the region I\hspace{-.1em}I $(0<\theta\leq \pi)$, respectively. 
For each patch, gauge fields are well-defined.
This gauge field configuration is nothing but the Dirac monopole
with the monopole charge $q$.
In the overlap region, the configurations on the region I 
and the region I\hspace{-.1em}I are related by the gauge transformation
\begin{align}
U_{\mathrm{I\hspace{-.1em}I}\rightarrow \mathrm{I}}
=\exp\left\{i\frac{4\pi}{\mu k} |v|^2
\cdot \varphi
\right\}
=\exp\left\{
i \: 2q\varphi
\right\},
\end{align}
which is single value since $q\in \bm{Z}/2$.

As discussed in \cite{Aharony:2008ug}, even after gauge fixing ABJM theory, 
there is a discrete 
redundant gauge symmetry left, which results in the following identification of scalar fields:
\begin{align}
Y^{A}\sim e^{2\pi i/k}Y^{A}.
\label{orbifold}
\end{align}

For the $1/2$-BPS solutions (\ref{1/2 BPS scalar field}) and (\ref{1/2 BPS gauge field}), 
we can calculate the energy $E$ and the $R$-charge $J_4$ 
(the charge corresponding to the rotation of the phase of $Y^4$);
\begin{align}
E
&=\int \frac{d\Omega}{\mu^2} 
\left(|\partial_t Y^A|^2+|\nabla_{a'}Y^A|^2+\frac{\mu^2}{4}|Y^A|^2\right)
= \mu k q, \n
J_4
&=\int \frac{d\Omega}{\mu^2} \left(
-iY^4 \partial_t Y_4^\dagger +i\partial_t Y^4 Y_4^\dagger
\right)
= 2k q, 
\end{align}
where $a'=1,2$.
Note that the solution saturates the following BPS bound\footnote{
The $\frac{1}{2}$ in the right-hand side is due to our $R$-charge assignment.
} 
\begin{align}
E=\frac{\mu}{2}J_4.
\label{bps}
\end{align}

\subsection{$1/4$-BPS solution}

Next, we will find 1/4-BPS solutions. 
In addition to the $1/2$-BPS condition \eqref{1/2 BPS condition}  
we further impose the following conditions
\begin{align}
i\gamma^0 \eta^{(+)}_{A'4} &=  \eta^{(+)}_{A'4}, \n
i\gamma^0 \eta^{(-)}_{A'B'} &= - \eta^{(-)}_{A'B'},
\label{1/4 BPS condition}
\end{align}
where the second condition is the complex conjugate of the first, 
so this gives rise to a $1/4$-BPS condition. 
In this case, \eqref{xi in terms of eta} becomes
\begin{align}
\xi^{(+)}_{A'4}
&= e^{i\frac{\mu t}{2}}e^{-i \frac{\phi}{2}}
\left(\cos\frac{\theta}{2}+\gamma^1 \sin\frac{\theta}{2}\right) \eta^{(+)}_{A'4}, \n
\xi^{(-)}_{A'B'}
&= e^{-i\frac{\mu t}{2}}e^{i \frac{\phi}{2}}
\left(\cos\frac{\theta}{2}+\gamma^1 \sin\frac{\theta}{2}\right) \eta^{(-)}_{A'B'}.
\end{align}
Substituting this into \eqref{del psi U(1)}, we obtain the following conditions for the scalars
\begin{align}
&Y^1=Y^2=Y^3=0, \n
&\partial_t Y^4+i\frac{\mu}{2} Y^4 -\mu \partial_\varphi Y^4=0, \n
&\partial_\theta Y^4+i\cot\theta \partial_\varphi Y^4=0.
\label{1/4 BPS scalar field}
\end{align}
It is easily seen that 
$Y^4\sim \sin^p\theta e^{ip\varphi}e^{-i(p+\frac{1}{2}) \mu t}$ solves the above equation
as well as the equation of motion.
So the general solution of the scalar fields is given by 
\begin{align}
&Y^1=Y^2=Y^3=0, \n
&Y^4=\sum_{p\in \bm{Z}_{\geq 0}+\frac{n}{k}} 
v_p \sin^p\theta e^{ip\varphi}e^{-i(p+\frac{1}{2}) \mu t},
\label{1/4 BPS scalar}
\end{align}
where $n$ is an integer in the range of $0\leq n \leq k-1$ and
$v_p$ are complex constants. 
When $p$ is an integer, $\sin^p\theta e^{ip\varphi}$ is 
the spherical Harmonics of $l=m=p$, $Y_{pp}(\theta,\varphi)$.
Here we have chosen $p$ 
in such a way that the solution is regular at $\theta=0, \pi$ and 
single-valued with \eqref{orbifold} under the shift $\varphi\rightarrow \varphi+2\pi$.
As in the $1/2$-BPS case, 
the $1/4$-BPS solution \eqref{1/4 BPS scalar} breaks $SU(4)$ $R$-symmetry to $SU(3)$.
From the equations of motion of the gauge fields in \eqref{EOM},
one can compute the gauge fluxes as
\begin{align}
F_{12}^{(1)}=F_{12}^{(2)}
&=\frac{2\pi \mu}{k}\sum_{p,p' \in \bm{Z}_{\geq 0}+\frac{n}{k}}(p+p'+1)
v_p(v_{p'})^* \sin^{p+p'}\theta
e^{i(p-p')(\varphi-\mu t)}, \n
F_{01}^{(1)}=F_{01}^{(2)}
&=\frac{2\pi \mu}{k}\sum_{p,p' \in \bm{Z}_{\geq 0}+\frac{n}{k}}(p+p')
v_p(v_{p'})^*\sin^{p+p'-1}\theta
e^{i(p-p')(\varphi-\mu t)}, \n
F_{02}^{(1)}=F_{02}^{(2)}
&=\frac{2\pi \mu i}{k}\sum_{p,p' \in \bm{Z}_{\geq 0}+\frac{n}{k}}
(p-p')v_p(v_{p'})^*\cos\theta\sin^{p+p'-1}\theta
e^{i(p-p')(\varphi-\mu t)}.
\label{1/4 BPS field strength}
\end{align}
Thus, in the general $1/4$-BPS solutions determined by 
\eqref{1/2 BPS condition} and \eqref{1/4 BPS condition},
in contrast to the $1/2$-BPS case,
not only $F_{12}^{(i)}$ but also $F_{0a'}^{(i)}$ ($a'=1,2$) are nonzero
and furthermore they have nontrivial $(t,\theta,\varphi)$ dependence. 
The quantization condition of the flux requires
\begin{align}
\frac{2\pi}{\mu k}\sum_{p\in\bm{Z}_{\geq 0}+\frac{n}{k}}
2^{2p+1} \frac{\Gamma(p+1)^2}{\Gamma(2p+1)} |v_p|^2=2q\in \bm{Z}_{\geq 0},
\label{1/4 BPS quantization condition}
\end{align}
where $q\in \bm{Z}_{\geq 0}/2$. 
So $v_p$ are given by
\begin{align}
v_p=\frac{e^{i\alpha_p}}{c_p}\sqrt{\frac{\mu k q_p}{2\pi}},
\label{v_p}
\end{align}
where 
\begin{align}
c_p=\sqrt{\frac{2^{2p} \Gamma(p+1)^2}{\Gamma(2p+1)}},
\label{def of cp}
\end{align}
$\alpha_{p}$ are real constants and 
$q_p$ are real constants with $\sum_p q_p=q$.
As in the $1/2$-BPS case, 
\eqref{1/4 BPS field strength} can be solved 
in terms of the gauge field with a gauge in which $A^{(1)}_1=A^{(2)}_1=0$ as
\begin{align}
&A^{(1)}_0=A^{(2)}_0 \n
&= \frac{2\pi}{k} \sum_{p\neq p' \in \bm{Z}_{\geq 0}+\frac{n}{k}} \!  (p+p')
v_p (v_{p'})^* e^{i(p-p')(\varphi-\mu t)} 
\sum_{r=0}^{\infty}
\frac{1}{2r+1}
\begin{pmatrix}
-\frac{p+p'}{2}+r \\
r
\end{pmatrix}
(\mp 1+\cos^{2r+1}\theta)
 \n
& \quad
+\frac{2\pi}{k} \sum_{p \in \bm{Z}_{\geq 0}+\frac{n}{k}} 2p|v_p|^2 
\sum_{r=0}^{\infty}
\frac{1}{2r+1}
\begin{pmatrix}
-p+r \\
r
\end{pmatrix}
\cos^{2r+1}\theta, \n
&A^{(1)}_1=A^{(2)}_1=0, \n
&A^{(1)}_2=A^{(2)}_2
=\frac{2\pi}{k} \: \sum_{p,p'  \in \bm{Z}_{\geq 0}+\frac{n}{k}} 
(p+p'+1) v_p (v_{p'})^*
e^{i(p-p')(\varphi-\mu t)}
\n &\hspace{5cm} \times
\frac{1}{\sin\theta} 
\sum_{r=0}^{\infty}\frac{1}{2r+1}
\begin{pmatrix}
-\frac{p+p'}{2}+r-1 \\
r
 \end{pmatrix}
(\pm 1-\cos^{2r+1}\theta),
\label{1/4 BPS gauge field}
\end{align}
	where $\begin{pmatrix}a\\b\end{pmatrix}$ is the binomial coefficient.
The upper and lower signs 
correspond to the region I $(0\leq \theta <\pi)$ 
and the region I\hspace{-.1em}I $(0<\theta\leq \pi)$ on $S^2$, respectively. 
Since all components of the field strength are nonzero and take the nontrivial form,
in the present gauge, not only $A_2^{(i)}$ but also $A_0^{(i)}$ are nonzero and 
 involve the $t$ and $\varphi$-dependence as well as the $\theta$-dependence.
(The $\theta$-dependence in $A_{2}^{(i)}$ 
seems to be a (higher order) generalization of the monopole configuration.)
The patch-dependence of $A_0^{(i)}$ is introduced 
so that $A_0^{(i)}$ does not have $\varphi$-dependence at $\theta=0$ and $\pi$.
Thus, on each patch, gauge fields are well-defined.
In the overlap region, 
one can transform the configurations of the gauge fields (\ref{1/4 BPS gauge field}) 
from one to the other by the transition function
\begin{align}
U_{\mathrm{I\hspace{-.1em}I}\rightarrow \mathrm{I}}
&=\exp\biggl\{
\frac{4\pi i}{\mu k} \!\!  \sum_{p\neq p' \in \bm{Z}_{\geq 0}+\frac{n}{k}} \!\!\!
2^{p+p'}\frac{\Gamma(\frac{p+p'}{2}+1)^2}{\Gamma(p+p'+1)}
 v_p (v_{p'})^*
\frac{e^{i(p-p')(\varphi-\mu t)}}{i(p-p')} 
+
2iq
\varphi
\biggr\}.
\end{align}
Note that
\begin{align}
\sum_{r=0}^{\infty}
\frac{1}{2r+1}
\begin{pmatrix}
-p+r-1 \\ r
\end{pmatrix}
&=\frac{2^{2p}\Gamma(p+1)^2}{\Gamma(2p+2)} \n
&=\frac{2p}{2p+1}
\sum_{r=0}^{\infty}
\frac{1}{2r+1}
\begin{pmatrix}
-p+r \\ r
\end{pmatrix}
\end{align}
The solution with $n=0$ and $v_l=0$ for $l\geq 1$ is the $1/2$-BPS solution 
discussed in the previous subsection.

Finally, we calculate charges for the $1/4$-BPS solutions. 
In addition to the energy and the $R$-charge computed in the $1/2$-BPS case, 
$1/4$-BPS solutions have nonzero momentum along $\varphi$ direction,
\begin{align}
E
&=\int \frac{d\Omega}{\mu^2} 
\left(|\partial_t Y^A|^2+|\nabla_{a'}Y^A|^2+\frac{\mu^2}{4}|Y^A|^2\right)
=2\pi \sum_{p\in\bm{Z}_{\geq 0}+\frac{n}{k}} (2p+1)c_p^2|v_p|^2, \n
J_4
&=\int \frac{d\Omega}{\mu^2} \left(
-iY^4 \partial_t Y_4^\dagger +i\partial_t Y^4 Y_4^\dagger
\right) 
=2k q, \n
P_\varphi
&=\int \frac{d\Omega}{\mu^2} \left(
-\partial_t Y^A \partial_\varphi Y_A^{\dagger}+\partial_\varphi Y^{A}\partial_t Y_A^\dagger
\right)
=\frac{2\pi}{\mu} \sum_{p\in\bm{Z}_{\geq 0}+\frac{n}{k}}
2p c_p^2|v_p|^2.
\end{align}
So the $1/4$-BPS solution satisfies the following BPS bound
\begin{align}
E=\mu \left(\frac{1}{2}J_4+P_{\varphi}\right).
\end{align}

\section{SYM on $\bR\times \bS^2$ from ABJM on $\bR\times \bS^2$}
\setcounter{equation}{0}

In this section we ``Higgs'' ABJM theory on $\bR\times \bS^2$ 
around a $1/2$-BPS solution following the procedure first discussed in \cite{Mukhi:2008ux}. 
In \cite{Mukhi:2008ux}, Mukhi and Papageorgakis had shown that one can obtain $\cN=8$ SYM from 
BLG theory on $\bR^3$ by expanding it around a vacuum $Y^A=\delta^{A4}v \bm{1}_N$ 
and taking the limit in which $v\rightarrow \infty$ and $k\rightarrow \infty$ 
with $v^2/k$ fixed. This procedure was called the ``novel Higgs mechanism''.

Here we will show that when a similar procedure is carried out around a $1/2$-BPS solution 
in ABJM theory on $\bR\times \bS^2$, the action reduces to $\cN=8$ SYM on $\bR\times \bS^2$,
which has interesting features such as the existence of many discrete vacua,
 a mass gap and $SU(2|4)$ symmetry (16 supercharges)\footnote{In the abelian case, 
the relation between the theory of a single M2-brane and the abelian SYM on $\bR\times \bS^2$
has been discussed in \cite{Maldacena:2002rb}.}.
Some details of $\cN=8$ SYM on $\bR\times \bS^2$ are summarized in appendix C.
Since $\cN=8$ SYM in three dimensions is not conformal,  
the theory on $\bR\times \bS^2$ is not related to that on $\bR^3$ in any simple way, 
unlike ABJM theory.
It should be noted that
the theory expanded around a $1/2$-BPS solution of ABJM theory on $\bR\times \bS^2$ 
has 12 supersymmetries and $SU(3)$ $R$-symmetry 
while $\cN=8$ SYM on $\bR\times \bS^2$ has 16 supersymmetries and $SU(4)$ $R$-symmetry,
so in the Higgsing we will see the enhancement of the $R$-symmetry 
as well as the number of supersymmetries.

\subsection{$\cN=8$ SYM on $\bR\times \bS^2$ around trivial vacuum}

We first consider $U(N)\times U(N)$ ABJM theory on $\bR\times \bS^2$
and expand it around the following $1/2$-BPS background, 
which is proportional to unit matrix:
\begin{align}
&Y^1=Y^2=Y^3=0, \quad Y^4=v e^{-i\frac{\mu t}{2}}\cdot \bm{1}, \n
&A^{(1)}_0=A^{(2)}_0=0, \quad A^{(1)}_1=A^{(2)}_1=0, \n
&A^{(1)}_2=A^{(2)}_2
=\frac{2\pi v^2}{k} \frac{\pm 1 - \cos\theta}{\sin\theta}\cdot \bm{1},
\label{bg for SYM around the trivial vacuum}
\end{align}
where $v=\sqrt{\frac{\mu k}{2\pi}q}$.
We have chosen $v$ to be real by using $U(1)_b$ symmetry.
We expand the fields in (\ref{ABJM}) around \eqref{bg for SYM around the trivial vacuum} as
\begin{align} 
&Y^A\rightarrow \hat{Y}^A+Y^A, \quad A^{(1)}\rightarrow \hat{A}^{(1)}+A^{(1)}, \quad
A^{(2)}\rightarrow \hat{A}^{(2)}+A^{(2)},
\label{expansion around the bg}
\end{align}
where the hat denotes the background.
The limit in which the ABJM theory reduces to SYM is
\begin{align}
q\rightarrow \infty \quad \text{and} \quad k\rightarrow \infty 
\quad \text{with} \quad \frac{4\pi \mu q}{k}=\frac{8\pi^2 v^2}{k^2}\equiv g^2 \quad 
\text{fixed},
\label{Higgsing limit}
\end{align}
where $g$ will be identified with the gauge coupling of $\cN=8$ SYM on $\bR\times \bS^2$ 
shortly\footnote{The fact that $g^2$ is identified with $\frac{8\pi^2v^2}{k^2}$ 
instead of $\frac{2\pi v^2}{k}$ as in the BLG case is a matter of notation, 
and one can go from one to the other by scaling fields by appropriate factors of $k$.}.
In this limit, the backgrounds $\hat{Y}^4$, $\hat{A}^{(1)}$ and $\hat{A}^{(2)}$ are $\cO(k)$.
To proceed with the computation, it is convenient to rewrite the gauge fields as follows
\begin{align}
A^{(1)}_m&=A_m+\frac{1}{2k}B_m, \n
A^{(2)}_m&=A_m-\frac{1}{2k}B_m.
\label{redef of A}
\end{align}
It turns out that in the limit (\ref{Higgsing limit})
$B_m$ becomes auxiliary fields and can be integrated out
while $A_m$ becomes dynamical and will be identified with the gauge field of SYM.

\subsubsection*{bosonic part}

Ignoring the terms of $\cO(k^{-1})$, we obtain
\begin{align}
&\int dt \frac{d\Omega}{\mu^2} \Tr\biggl[
-|D'_{a}Y^{A'}|^2 -\frac{\mu^2}{4}Y^{A'}Y_{A'}^\dagger
+|D'_0Y^4+\frac{i}{k}\hat{Y}^4B_0|^2 
-\frac{\mu}{2k}(\hat{Y}^4Y_4^\dagger+\hat{Y}_4^\dagger Y^4)B_0 
\n
&\qquad
-|D'_1Y^4+\frac{i}{k}\hat{Y}^4B_1|^2
-|D'_2Y^4+\frac{i}{k}\hat{Y}^4B_2|^2
-\frac{\mu^2}{4}Y^4Y_4^\dagger
+\frac{1}{2\pi}(B_0F_{12}+B_1F_{20}+B_2F_{01}) 
\n
&\qquad
+\frac{4\pi^2}{k^2}|\hat{Y}^4|^2\left(
[Y_{A'}^\dagger,Y^{B'}][Y^{A'},Y_{B'}^\dagger]
+[Y^{A'},Y^{B'}][Y_{A'}^\dagger,Y_{B'}^\dagger] 
\right)
+\frac{8\pi^2}{k^2}|\hat{Y}^4|^2 [\phi,Y^{A'}][\phi,Y_{A'}^\dagger]
\biggr],
\label{Higgsing 1}
\end{align}
where $D'_a=\nabla_{a}+i[A_{a},\cdot]$.
Integrating out $B_a$ and rewriting $Y^{A'}$ ($A'=1,2,3$) and $Y^4$ as
\begin{align}
Y^{A'}&=\frac{1}{\sqrt{2}g} X^{A'4}, \n
Y_{A'}^\dagger&=\frac{1}{\sqrt{2}g}X_{A'4}
=\frac{1}{\sqrt{2}g}\cdot\frac{1}{2}\epsilon_{A'B'C'}X^{B'C'}, \n
Y^4&=\frac{e^{-i\frac{\mu t}{2}}}{\sqrt{2}g}(\phi+i\rho),
\label{redef of Y}
\end{align}
we finally get
\begin{align}
&\frac{1}{g^2}\int dt \frac{d\Omega}{\mu^2} \Tr\biggl[-\frac{1}{2}D'_m\phi{D'}^m\phi 
-\frac{1}{2}\left(F_{12}-\mu \phi\right)^2
+\frac{1}{2}(F_{01})^2
+\frac{1}{2}(F_{20})^2 \n
&\quad
-\frac{1}{2}D'_m X_{AB} {D'}^m X^{AB}-\frac{\mu^2}{8}X_{AB}X^{AB} 
+\frac{1}{4} [X_{AB},X_{CD}][X^{AB},X^{CD}]
+\frac{1}{2} [\phi,X_{AB}][\phi,X^{AB}]
\biggr].
\label{SYM on RxS2 bosonic part}
\end{align}
To obtain this expression, we have integrated by parts 
and used Bianchi identity $\epsilon^{abc}D'_aF_{bc}=0$.
The action \eqref{SYM on RxS2 bosonic part} is invariant under $U(N)$ gauge transformation,
where the scalar fields $\phi$ and $X_{AB}$ transform as the adjoint representation of 
$U(N)$ and $D'_m$ is the adjoint covariant derivative with the gauge field $A_m$,
and also has global $SU(4)$ symmetry.
This theory is nothing but (the bosonic part of) $\cN=8$ SYM on $\bR\times \bS^2$.

\subsubsection*{fermionic part}

The details of the fermionic part of $\cN=8$ SYM action are also reproduced by this 
procedure. The fermionic part of ABJM action has two set of terms: 
the kinetic term as well as the quartic interaction term involving the fermions 
and bosons. 
It turns out from \eqref{redef of A} that
the effect of the Higgsing procedure on the covariant derivative for the fermions 
is simply to drop the $B_{m}$ field in the covariant derivative of ABJM action
\begin{equation}
D_{m}\psi_{A}\rightarrow D'_{m}\psi_{A}
=\nabla_{m}\psi_{A} + i[A_{m},\psi_{A}],
\end{equation}
Then the kinetic term of ABJM theory becomes
\begin{align}
\Tr\left(i\psi^{\dagger A}\gamma^{m}D'_{m}\psi_{A}\right). 
\label{kinetic term after higgsing}
\end{align}
Note that $\psi_A$ here is the fermion field of the SYM and becomes adjoint field in $U(N)$.
We now come to the quartic terms, the last two lines in \eqref{ABJM}.
By the Higgsing those terms reduce to
\begin{align}
\Tr\biggl(
&2ie^{i\frac{\mu t}{2}}\psi^{\dagger 4}[X^{4A'},\psi_{A'}]
-2ie^{-i\frac{\mu t}{2}}\psi_4[X_{4A'},\psi^{\dagger A'}]
+i\psi^{\dagger A'}[\phi,\psi_{A'}]
-i\psi^{\dagger 4}[\phi,\psi_4] \n
&-ie^{-i\frac{\mu t}{2}}\psi^{\dagger A'}[X_{A'B'},\psi^{\dagger B'}]
+ie^{i\frac{\mu t}{2}}\psi_{A'}[X^{A'B'},\psi_{B'}]
\biggr),
\label{qurtic terms after higgsing}
\end{align}
where $X^{AB}$ are defined in \eqref{redef of Y}.

In what follows, we see that these two, \eqref{kinetic term after higgsing}
and \eqref{qurtic terms after higgsing}, can be rewritten in $SU(4)$ symmetric form 
and are indeed the fermionic part of $\cN=8$ SYM.
First we absorb the time-dependence appearing in \eqref{qurtic terms after higgsing}
by the following redefinition
\begin{align}
\psi_{A'}&\rightarrow e^{-i\frac{\mu t}{4}}\psi_{A'}, \n
\psi_{4}&\rightarrow e^{i\frac{\mu t}{4}}\psi_{4}.
\end{align}
By this, the kinetic term yields mass terms
\begin{align}
\Tr\left(i\psi^{\dagger A}\gamma^{m}D'_{m}\psi_{A}\right)
\rightarrow 
\Tr\left(
i\psi^{\dagger A}\gamma^{m}D'_{m}\psi_{A}
+\frac{\mu}{4}\psi^{\dagger A'}\gamma^0 \psi_{A'}
-\frac{\mu}{4}\psi^{\dagger 4}\gamma^0 \psi_{4}
\right).
\label{kinetic term after higgsing 2}
\end{align}
Next, in order to see the $SU(4)$ invariance of the action,
we regard $\psi_4$ ($\psi^{\dagger 4}$) which transforms 
as the forth-component of $\bm{4}$ ($\bar{\bm{4}}$) of $SU(4)$ in ABJM theory
as the field which transforms as the forth-component of $\bar{\bm{4}}$ ($\bm{4}$).
Namely, we interchange $\psi_4$ and $\psi^{\dagger 4}$;
\begin{align}
\psi_4 \leftrightarrow \psi^{\dagger 4}.
\label{interchange}
\end{align}
The reason of this interchange is explained below.
Then \eqref{qurtic terms after higgsing} and \eqref{kinetic term after higgsing 2} 
are rewritten in $SU(4)$ symmetric form as
\begin{align}
\Tr\biggl(
i\psi^{\dagger A}\gamma^{m}D'_{m}\psi_{A}
+\frac{\mu}{4}\psi^{\dagger A}\gamma^0 \psi_{A} 
+i\psi^{\dagger A}[\phi,\psi_{A}]
-i\psi^{\dagger A}[X_{AB},\psi^{\dagger B}]
+i\psi_{A}[X^{AB},\psi_{B}]
\biggr)
\end{align}
The precise correspondence with the form of $\cN=8$ SYM on $\bR\times \bS^{2}$ given 
in appendix C
can be seen by performing the following replacements:
$\mu\rightarrow -\mu$, $\phi\rightarrow -\phi$,
$\psi_A\rightarrow \gamma^0\hat{\psi}^\dagger _A$ and
$\psi^{\dagger A}\rightarrow \gamma^0\hat{\psi}^{A}$,
where $\hat{\psi}^A$ and $\hat{\psi}_A^\dagger$ are fermions of $\cN=8$ SYM.

The fermions of ABJM theory $\psi_A$ and $\psi^{\dagger A}$ transform 
as $\bm{4}_{1}$ and $\bar{\bm{4}}_{-1}$ 
under $SU(4)\times U(1)_b$, respectively.  
By the Higgsing mechanism, 
$SU(4)$ is broken into $SU(3)\times U(1)$, and thus
$\psi_A$ and $\psi^{\dagger A}$ are split into 
$\bm{3}_{1/2}\oplus \bm{1}_{3/2}$ and
$\bar{\bm{3}}_{-1/2}\oplus\bm{1}_{-3/2}$, respectively.
On the other hand, 
the fermions of $\cN=8$ SYM are $\bm{4}$ and $\bar{\bm{4}}$ of $SU(4)$ 
and not charged under $U(1)_b$ since they are adjoint fields.
By decomposing $SU(4)$ into $SU(3) \times U(1)$, 
$\hat{\psi}_A^\dagger$ and $\hat{\psi}^A$
are split into $\bm{3}_{1/2}\oplus\bm{1}_{-3/2}$ 
and $\bar{\bm{3}}_{-1/2}\oplus\bm{1}_{3/2}$, respectively.
To identify the fermions of the ABJM theory with those of $\cN=8$ SYM,
we have to set $\psi_{A'}=\hat\psi^{\dagger}_{A'}$ 
and $\psi_4=\hat\psi^{4}$ essentially.
This is what we have done in the above.
 (See details in appendix D).
\\

Note that the scalar field $\rho$, 
which is the fluctuation of $Y^4$,
 is completely decoupled from the theory 
since in the limit \eqref{Higgsing limit} 
$\rho$ becomes a compact scalar with period $\rho\sim \rho + g^2$,
which can be seen from the identification of scalars \eqref{orbifold} 
with \eqref{bg for SYM around the trivial vacuum}, 
\eqref{expansion around the bg}, \eqref{Higgsing limit} and \eqref{redef of Y}.
Note also the difference of the action of $\cN=8$ SYM on $\bR\times \bS^2$ 
from that on the flat space.
For instance,
the scalar field $\phi$ has the different mass from that of other scalars 
and the coupling with $F_{12}$ and so 
there is no $SO(7)$ global symmetry among scalar fields unlike $\cN=8$ SYM on $\bR^{1,2}$ 
where there is no such difference among scalar fields and the $SO(7)$ global symmetry exists.
From the perspective of the Higgsing,
the scalar field $\phi$ is coming from the fluctuation around 
the $1/2$-BPS solution \eqref{1/2 BPS scalar field} of $Y^4$ as \eqref{redef of Y} and
the difference from other scalars is coming from the time-dependence of 
the background around which we expanded ABJM theory on $\bR\times \bS^{2}$. 
This time-dependence is also the source of the mass term of the fermions in the SYM. 
Now, $\cN=8$ SYM on $\bR\times \bS^2$ can also be obtained from the dimensional 
reduction of $\cN=4$ SYM on $\bR\times \bS^3(/\mathbb{Z}_n)$
onto $\bR\times \bS^2$, where $\bS^3$ is viewed as $\bS^1$ fiber over $\bS^2$ \cite{Lin:2005nh}. 
It is interesting to note the different origin of the scalar field $\phi$ and 
the mass terms from this viewpoint. In this construction, the scalar field 
$\phi$ in $\cN=8$ SYM on $\bR\times \bS^2$ originates from the gauge field along 
the fiber direction in $\cN=4$ SYM on $\bR\times \bS^3(/\mathbb{Z}_n)$ 
and the mass term of the scalar $\phi$ 
and that of the fermions
from the difference of the spin connection of $\bS^3$ and $\bS^2$.

One can also carry out the higgsing procedure directly at the level of 
the supersymmetry transformations of ABJM theory and  show that 
it reduces to a subset of the full supersymmetry transformations of the SYM\footnote{
In \cite{Fujimori:2010ec}, 
the BPS equations of ABJM theory on flat space was shown to reduce to the BPS equations of 
SYM under Higgsing.}.
The supersymmetry transformation of ABJM theory \eqref{SUSY transf} reduces to 
that of $\cN=8$ SYM \eqref{SUSY transformation of SYM} by
\begin{equation}
\frac{i}{\sqrt{2}} e^{-i\mu t/4} \xi_{4B'}^{(+)} = \varepsilon^\dagger_{B'}, \quad
\frac{i}{\sqrt{2}} e^{i\mu t/4} \xi^{(+)4B'} = -\varepsilon^{B'}, \quad
\end{equation}
with $\varepsilon^4, \varepsilon^\dagger_4=0$. 
This means that 
the enhanced supersymmetry is given by $\varepsilon^4, \varepsilon^\dagger_4$. 
We will now briefly 
comment on the symmetry enhancement that happens during the Higgsing process.

While $\cN=8$ SYM on $\bR\times \bS^2$ as well as on flat space 
preserves sixteen supersymmetries, the half-BPS solution of ABJM theory, 
around which the Higgsing takes place, preserves only twelve supersymmetries. 
Therefore the Higgsing procedure is 
accompanied with an enhancement of supersymmetry as well as an enhancement of the associated 
R-symmetry. This is different from  the case of higgsing in the BLG theory, where there is no 
enhancement of symmetry, since the vacuum of the BLG theory preserves sixteen supersymmetries 
to begin with.

There is a simple way to understand how this enhancement happens during the process of Higgsing.
The effect of the Higgsing can be summarized by some ``effective higgsing rules'', 
as was done for the BLG case\cite{Ezhuthachan:2009sr}.
In particular, under the Higgsing procedure, 
the bi-fundamental covariant derivative action on fields $Y^{A'}$, $Y^{\dagger}_{A'}$ 
($A'=1,2,3$) $(D_{m}Y^{A'}=\partial_{m}Y^{A'} +iA^{(1)}_{m}Y^{A'} -iY^{A'} A^{(2)}_{m})$ 
is replaced by an adjoint covariant derivative: 
$(D'_{m}Y^{A'} =\partial_{m}Y^{A'} +i[A_{m},Y^{A'}])$.  
This is true for the covariant derivative of the fermions as well. 
The solution around which the Higgsing is done preserves only 
$SU(3)\times U(1)$  of the full global symmetry $SU(4)\times U(1)_b$ of ABJM theory. 
The conserved currents associated with these symmetries are gauge invariant observables 
constructed of the $Y^{A'}$ and the $Y^{\dagger}_{A'}$  and take the form: 
\begin{equation}
J^{A'}_{B' m} =  \textrm{Tr}(Y^{A'}D_{m}Y^{\dagger}_{B'})
\end{equation}
The conserved currents associated to the $SO(6)$ symmetry of the SYM would be :
\begin{equation}
j^{A'}_{B' m} = \textrm{Tr}(Y^{A'}D'_{m}Y^{\dagger}_{B'}); \; \; 
\hat{j}^{A'B'}_{m} = \textrm{Tr}(Y^{[A'}D'_{m}Y^{B']}); \; \; 
\hat{j}^{\dagger}_{A'B' m} = \textrm{Tr}(Y^{\dagger}_{[A'}D'_{m}Y^{\dagger}_{B']})
\end{equation}
 The additional currents which arise in the SYM limit descend from operators which were not 
gauge invariant observables in ABJM theory. 
They become gauge invariant, after Higgsing, 
under the gauge transformations of the reduced gauge group. 
This discussion carries over to the enhancement of supercurrents as well.

\subsection{$\cN=8$ SYM on $\bR\times \bS^2$ around nontrivial vacua}

We can also obtain $\cN=8$ SYM on $\bR\times \bS^2$ expanded around a nontrivial vacuum,
which is presented in appendix C.
To see this, let us choose a more general $1/2$-BPS background,  
which is diagonal but not proportional to unit matrix;
\begin{align}
&Y^1=Y^2=Y^3=0, 
 \quad Y^4=\diag\left(v_{1},v_2,\cdots,v_N\right) e^{-i\frac{\mu t}{2}}, \n
&A^{(1)}_0=A^{(2)}_0=0, \quad A^{(1)}_1=A^{(2)}_1=0, \n
&A^{(1)}_2=A^{(2)}_2
= \frac{2\pi}{k}|Y^4|^2 \frac{\pm 1 - \cos\theta}{\sin\theta}.
\label{bg for SYM around the nontrivial vacuum}
\end{align}
Here
\begin{align}
v_{i}=\sqrt{\frac{\mu k}{2\pi}(q+q_i)},
\end{align}
where $q$ and $q_{i}$ are positive half-integers.
The theory expanded around such a background is equivalent to 
the one expanded around \eqref{bg for SYM around the trivial vacuum} 
in which the fluctuation of $Y^4$, for instance, is replaced by
\begin{align}
(Y^4)_{ij}\rightarrow 
(Y^4)_{ij}+\delta_{ij}(v_{i}-v)e^{-i\frac{\mu t}{2}}.
\end{align}
In the limit \eqref{Higgsing limit}, $v_i-v$ becomes
\begin{align}
v_i-v \rightarrow \frac{\mu}{\sqrt{2}g}q_{i} 
\end{align}
and so is regarded as the background of the fluctuation.
Under the Higgsing around \eqref{bg for SYM around the nontrivial vacuum}, 
ABJM theory on $\bR\times \bS^2$, therefore, 
reduces to $\cN=8$ SYM on $\bR\times \bS^2$ expanded around
\begin{align}
\phi&=\mu \diag(q_1,q_2,\cdots,q_N), \quad X_{AB}=0, \n
A_0&=0, \quad A_1=0, \quad
A_2
=\phi \frac{\pm 1 - \cos\theta}{\sin\theta}.
\label{nontrivial vacuum of SYM}
\end{align}
Since the solution \eqref{bg for SYM around the nontrivial vacuum} 
we expanded the ABJM theory around
is also $1/2$-BPS as in the previous case, 
it is expected that \eqref{nontrivial vacuum of SYM} keeps same amount of supersymmetries as
the trivial vacuum of $\cN=8$ SYM on $\bR\times \bS^2$.
Indeed, as presented in appendix C
the configuration \eqref{nontrivial vacuum of SYM} is a (nontrivial) vacuum of 
$\cN=8$ SYM on $\bR\times \bS^2$.

\subsection{$\cN=8$ SYM on $\bR\times \bS^2$ around $1/2$-BPS solution}

It is also possible to obtain $\cN=8$ SYM on $\bR\times \bS^2$ expanded around 
$1/2$-BPS solutions by Higgsing ABJM theory on $\bR\times \bS^2$ 
about a diagonal $1/4$-BPS solution in which $Y^A$ take the form
\begin{align}
&Y^1=Y^2=Y^3=0, \n 
&(Y^4)_{ij}=\delta_{ij}\sum_{p\in\bm{Z}_{\geq 0}+\frac{n}{k}} 
v_{ip} \sin^p\theta e^{ip\varphi-i(p + \frac{1}{2})\mu t}.
\label{1/4 BPS scalar for Higgsing}
\end{align}
In particular, we first take a solution with $n=0$, namely $p=l\in \bZ_{\geq 0}$.
The gauge field configuration is also diagonal and 
each component is given by \eqref{1/4 BPS gauge field} 
with $v_p$ replaced by $v_{il}$ for each component. 
In particular, we choose $v_{il}$ as
\begin{align}
v_{i0}&=
\sqrt{\frac{\mu k}{2\pi}(q+q_{i0}+\beta_{i0})}, \n
v_{il}&=\frac{e^{i\alpha_{il}}}{c_l}\sqrt{\frac{\mu k}{2\pi}\beta_{il}} \;\; (l\geq 1),
\label{v to obtain 1/2 BPS solution}
\end{align}
where $q$ and $q_{i0}$ are positive half-integers and $\beta_{il}$ are real constants with
$\sum_{l\geq 0}\beta_{il}=0$. $c_l$ is defined in \eqref{def of cp} and $\alpha_{il}$
are real constants.
ABJM theory  around this background is the same as the one around the background 
\eqref{bg for SYM around the trivial vacuum} with the fluctuation of $Y^4$ replaced
by
\begin{align}
(Y^4)_{ij}\rightarrow (Y^4)_{ij}+\delta_{ij}\Big(
\sum_{l\geq 0} v_{il} \sin^l\theta e^{il\varphi-i(l + \frac{1}{2})\mu t}
-v e^{-i\frac{\mu t}{2}} \Big).
\label{replacement of Y^4 2}
\end{align}
Then, under the limit in which
\begin{align}
&q\rightarrow \infty, \quad  k\rightarrow \infty \quad \text{and} \quad 
\beta_{il}\rightarrow 0
\quad \text{with} \quad 
\frac{4\pi \mu q}{k}\equiv g^2 
\quad \text{and} \quad 
v_{il}(\sim \sqrt{k \beta_{il}})  \quad \text{fixed}. 
\label{Higgsing limit 2}
\end{align}
the second term in the right-hand side in \eqref{replacement of Y^4 2} becomes
\begin{align}
\sum_{l\geq 0} 
&v_{il} \sin^l\theta e^{il\varphi-i(l + \frac{1}{2})\mu t}-v e^{-i\frac{\mu t}{2}} \n
&\rightarrow 
\frac{\mu}{\sqrt{2}g}q_{i0}
e^{-i\frac{\mu t}{2}}
+
\sum_{l\geq 1}v_{il}\sin^l\theta e^{il\varphi-i(l + \frac{1}{2})\mu t},
\end{align}
So, the theory we finally get is $\cN=8$ SYM on $\bR\times \bS^2$ around
\begin{align}
\phi_{ij}
&=\delta_{ij}\Bigl(\mu q_{i0}
+\frac{g}{\sqrt{2}}\sum_{l\geq 1}
\sin^l\theta(v_{il}e^{il(\varphi-\mu t)}+\text{c.c.})\Bigr), \n
X_{AB}&=0, \n
(A_0)_{ij}
&=\delta_{ij} \: 
\frac{g}{\sqrt{2}} \sum_{l\geq 1 } \!  l (v_{il}e^{il(\varphi-\mu t)}+\text{c.c.})
\sum_{r=0}^{\infty}
\frac{1}{2r+1}
\begin{pmatrix}
-l+r \\
r
\end{pmatrix}
(\mp 1+\cos^{2r+1}\theta), \n
A_1&=0, \n
(A_2)_{ij}
&=\delta_{ij}\bigg[\mu q_{i0} \frac{\pm 1 - \cos\theta}{\sin\theta}
\n &\hspace{1cm}
+\frac{g}{\sqrt{2}} \: \sum_{l\geq 1} 
(l+1) (v_{il}e^{il(\varphi-\mu t)}+\text{c.c.})
\n &\hspace{4cm}\times
\frac{1}{\sin\theta} 
\sum_{r=0}^{\infty}\frac{1}{2r+1}
\begin{pmatrix}
-l+r-1 \\
r
 \end{pmatrix}
(\pm 1-\cos^{2r+1}\theta)\biggr].
\label{1/2 BPS solution of SYM}
\end{align}
The field strength for the above gauge field configuration is give by
\begin{align}
(F_{01})_{ij}
&=\delta_{ij}\frac{\mu g}{\sqrt{2}}\sum_{l\geq 1} 
l \sin^{l-1}\theta \left(v_{il} e^{il(\varphi-\mu t)}+\text{c.c.}\right), \n
(F_{02})_{ij}
&=\delta_{ij}\frac{\mu g i}{\sqrt{2}}\sum_{l\geq 1}
l\cos\theta\sin^{l-1}\theta \left(v_{il} e^{il(\varphi-\mu t)}-\text{c.c.}\right), \n
(F_{12})_{ij}
&=\delta_{ij}\left(\mu^2q_{i0}+\frac{\mu g}{\sqrt{2}}\sum_{l\geq 1}
(l+1)\sin^l\theta \left(v_{il} e^{il(\varphi-\mu t)}+\text{c.c.}\right)\right).
\end{align}
It turns out from the Killing spinor equation $\delta \hat{\psi}^A=0$ of 
$\cN=8$ SYM on $\bR\times \bS^2$ given in appendix C 
that the field configuration \eqref{1/2 BPS solution of SYM} 
is a $1/2$-BPS solution of the SYM\footnote{
As discussed in \cite{Maldacena:2002rb} (also in \cite{Ishiki:2006yr}), 
the plane wave (BMN) matrix model can be regarded as a matrix regularization of 
$\cN=8$ SYM on $\bR\times \bS^2$.
So, there should be $1/2$-BPS solutions in the plane wave matrix model 
corresponding to \eqref{1/2 BPS solution of SYM}. 
Indeed one of $1/2$-BPS solutions in the plane wave matrix model studied in \cite{Hoppe:2007tv} 
seems to correspond to \eqref{1/2 BPS solution of SYM}.}.

One can also carry out the Higgsing
to a solution with $n\neq 0$ in \eqref{1/4 BPS scalar for Higgsing}.
In the same manner as before, we take $v_{ip}$ ($p\in \bZ_{\geq 0}+\frac{n}{k}$) as
\begin{align}
v_{i\frac{n}{k}}&=
\frac{1}{c_{\frac{n}{k}}}\sqrt{\frac{\mu k}{2\pi}(q+q_{i\frac{n}{k}}+\beta_{i\frac{n}{k}})}, \n
v_{ip}
&=\frac{e^{i\alpha_{ip}}}{c_{p}}
\sqrt{\frac{\mu k}{2\pi}\beta_{ip}} \;\; 
\left(p\in \bZ_{\geq 1}+\frac{n}{k}\right),
\end{align}
and take the limit in which
\begin{align}
&q\rightarrow \infty, \quad  k\rightarrow \infty \quad \text{and} \quad 
\beta_{ip}\rightarrow 0
\quad \text{with} \quad 
\frac{4\pi \mu q}{k}\equiv g^2 
\quad \text{and} \quad 
v_{ip}(\sim \sqrt{k \beta_{ip}})  \quad \text{fixed}. 
\end{align}
The effect of $n(\neq 0)$ results in extra terms being added to the previous result.
For instance, in the $k\rightarrow \infty$ limit,
$\sin^{\frac{n}{k}}\theta$ is approximated as
$\sin^{\frac{n}{k}}\theta
\rightarrow 1+\frac{n}{k}\ln \sin\theta+\cO((\frac{n}{k})^2)$,
which is valid except at $\theta = 0$ and $\pi$, 
and $v_{i(l+\frac{n}{k})}$ can be regarded as $v_{il}$ in \eqref{v to obtain 1/2 BPS solution}
times a constant:
\begin{align}
v_{i(l+\frac{n}{k})}
\rightarrow v_{il} 
\times  \left(1+\frac{n}{k}\ln 2 +\cO\Bigl(\Bigl(\frac{n}{k}\Bigr)^2\Bigr)\right).
\end{align}
Then, \eqref{1/4 BPS scalar for Higgsing} with $n\neq 0$ reduces to, 
except at $\theta = 0$ and $\pi$,
\begin{align}
&\sum_{p\in\bm{Z}_{\geq 0}+\frac{n}{k}} 
v_{ip} \sin^p\theta e^{ip\varphi-i(p + \frac{1}{2})\mu t} \n
&\rightarrow
v e^{-i\frac{\mu t}{2}}
+\left[
\frac{g}{2\sqrt{2}\pi}n\left(\ln\frac{\sin\theta}{2}+i(\varphi-\mu t)\right)
+\frac{\mu}{\sqrt{2}g} q_{i0}+\sum_{p\geq 1}v_p\sin^p\theta e^{ip(\varphi-\mu t)}\right]
e^{-i\frac{\mu t}{2}}.
\label{1/4 BPS Higgsing with nonzero n}
\end{align}
The second term is the new term arising due to the nonzero $n$.
One can easily carry out the same calculations for the gauge field configurations.
Thus the configurations in the SYM obtained from the $1/4$-BPS solutions with nonzero $n$ 
of ABJM theory via the Higgsing are
\begin{align}
\phi_{ij}
&=\delta_{ij}\Bigl(\mu q_{i0}+\frac{n g^2}{2\pi}\ln \frac{\sin\theta}{2}
+\frac{g}{\sqrt{2}}\sum_{l\geq 1}
\sin^l\theta(v_{il}e^{il(\varphi-\mu t)}+\text{c.c.})\Bigr), \n
X_{AB}&=0, \n
(A_0)_{ij}
&=\delta_{ij} \: \biggl[-\frac{\mu n g^2}{2\pi}\ln\tan\frac{\theta}{2} \n
&\quad
+\frac{g}{\sqrt{2}} \sum_{l\geq 1} \!  l (v_{il}e^{il(\varphi-\mu t)}+\text{c.c.})
\sum_{r=0}^{l-1
}
\frac{1}{2r+1}
\begin{pmatrix}
-l+r \\
r
\end{pmatrix}
(\mp 1+\cos^{2r+1}\theta)\biggr], \n
A_1&=0, \n
(A_2)_{ij}
&=\delta_{ij}\bigg[\mu q_{i0} \frac{\pm 1 - \cos\theta}{\sin\theta}
+\frac{ng^2}{2\pi}\left(\frac{1-\cos\theta}{\sin\theta}\ln\sin\frac{\theta}{2}
-\frac{1+\cos\theta}{\sin\theta}\ln\cos\frac{\theta}{2}\right)
\n &\hspace{1cm}
+\frac{g}{\sqrt{2}} \: \sum_{l\geq 1} 
(l+1) (v_{il}e^{il(\varphi-\mu t)}+\text{c.c.})
\n &\hspace{4cm}\times
\frac{1}{\sin\theta} 
\sum_{r=0}^{l
}
\frac{1}{2r+1}
\begin{pmatrix}
-l+r-1 \\
r
 \end{pmatrix}
(\pm 1-\cos^{2r+1}\theta)\biggr].
\end{align}
The field strength for the above gauge field configuration is give by
\begin{align}
(F_{01})_{ij}
&=\delta_{ij}\left(\frac{\mu n g^2}{2\pi}\frac{1}{\sin\theta}
+\frac{\mu g}{\sqrt{2}}\sum_{l\geq 1} 
l \sin^{l-1}\theta \left(v_{il} e^{ip(\varphi-\mu t)}+\text{c.c.}\right)\right), \n
(F_{02})_{ij}
&=\delta_{ij}\frac{\mu g i}{\sqrt{2}}\sum_{l\geq 1}
l\cos\theta\sin^{l-1}\theta \left(v_{il} e^{il(\varphi-\mu t)}-\text{c.c.}\right), \n
(F_{12})_{ij}
&=\delta_{ij}\left[\mu^2q_{i0}+\frac{\mu n g^2}{2\pi}\left(1+\ln\frac{\sin\theta}{2}\right)
+\frac{\mu g}{\sqrt{2}}\sum_{l\geq 1}
(l+1)\sin^l\theta \left(v_{il} e^{il(\varphi-\mu t)}+\text{c.c.}\right)\right].
\end{align}
Note that the terms proportional to $n$ appearing in $F_{01}$ and $A_{0}$
can be regarded as analogue on $\bR\times\bS^2$ of 
the Callan-Maldacena solution on flat space \cite{Callan:1997kz}, 
which describes a bound state of fundamental strings and D2-branes. 
This part in the solution represents 
$n$ fundamental strings attaching D2-branes 
on the north pole ($\theta=0$) and the south pole ($\theta=\pi$).  
The behavior around them indeed matches with the solution \cite{Kuroki:2011xn}.
On the other hand, 
the expressions for $F_{12}$ and $A_2$ are specific to the analysis on $\bR\times \bS^2$.
$F_{12}$ is singular at $\theta=0$ and $\theta=\pi$ but $A_{2}$ is not. 
Note also that the integral of the new term in $F_{12}$ 
 over $\bS^2$ vanishes as well as that of the terms of $l\geq 1$, 
so the flux quantization condition is just 
$\frac{1}{2\pi \mu^2}\int_{S^2}(F_{12})_{ii}=2q_{i0}\in \bm{Z}$,
which is consistent with that in ABJM theory.

\section{Summary and Discussion}
\setcounter{equation}{0}

In summary, we have solved BPS equations of ABJM theory on $\bR\times \bS^2$
for diagonal configurations
and shown that 
``Higgsing'' the ABJM theory around the $1/2$-BPS solution leads to  
$\mathcal{N}=8$ SYM on $\bR\times \bS^2$. 
The BPS solutions we found, in general, have nonzero angular momentum along $\varphi$ direction
and the non-trivial fluxes, not only $F_{12}$ but also $F_{01}$ and $F_{02}$. 
Higgsing around the $1/2$-BPS solution 
where the scalar field vev is proportional to the identity 
gives rise to $\mathcal{N}=8$ SYM on $\bR\times \bS^2$ expanded around the trivial vacuum 
while higgsing around $1/2$-BPS solutions 
which are diagonal but not proportional to the identity 
leads to the SYM expanded around a non-trivial vacuum. 
If we Higgs around a $1/4$-BPS configuration, 
then we end up getting the SYM expanded around a $1/2$-BPS solution. 
In fact, higgsing around various solutions of ABJM theory 
should reproduce the SYM expanded around its various solutions.  




Since the ABJM on $\bR\times \bS^2$ is dual to M-theory on global $AdS_{4}$, 
it is worth asking what the duals of the BPS solutions, we find in this paper, are. 
In \cite{Nishioka:2008ib}, Nishioka and Takayanagi solve 
the BPS equations explicitly in the bulk 
and construct a class of dual giant graviton solutions 
in M-theory on $AdS_{4}\times S^{7}/\mathbb{Z}_{k}$. 
In particular, they find a spinning dual giant graviton configuration.
The spinning dual giant graviton is a M2-brane expanding into $AdS_4$,
which rotates along the fiber coordinate of the $S^7$
($S^7$ being the fibration of $S^1$ over $\mathbb{C}\mathbb{P}^{3}$)
and spins along the azimuthal direction of $S^2\subset AdS_4$. 
This spinning dual giant graviton has a non-trivial profile along the $AdS_{4}$ 
and has been called the ``giant torus''. 
These solutions should be dual to the class of solutions we construct in this paper 
with nonzero $P_{\varphi}$ and $J_{4}$ 
corresponding to the nonzero spin and the angular momentum, respectively, in the bulk.

In a forthcoming paper \cite{ESY}, 
we will classify the space of solutions on the bulk side, 
which includes the giant torus solution, 
in terms of intersections of holomorphic surfaces with the target space, 
following \cite{Mikhailov:2000ya, Kim:2006he} 
and then using the methods given in \cite{Biswas:2006tj, Mandal:2006tk, Bhattacharyya:2007sa} 
we will compare and match with a similar classification 
on the space of boundary solutions presented here.
\vspace*{1ex}
\noindent{\bf Acknowledgment:} 
We would like to thank collectively 
Rajesh Gopakumar, Hikaru Kawai, Seok Kim, Tsunehide Kuroki, Suvrat Raju, Nemani Suryanarayana 
for useful discussions.  
The work of S.S. is supported in part by the JSPS Research Fellowship for Young Scientists.
SY is supported by National Research Foundation of Korea (NRF) grant No. 2010-0007512,
and No. 2009-0076297.

\appendix

\section{Conventions}
\setcounter{equation}{0}

In this paper, we consider the ABJM theory on $\bR\times \bS^2$ endowed with the metric
\begin{align}
ds^2=-dt^2+\frac{1}{\mu^2}\left(d\theta^2+\sin^2\theta d\varphi^2 \right),
\end{align}
where $\mu^{-1}$ is the radius of $S^2$. 
We take the local Lorentz frame as
\begin{align}
&e^0=dt,\quad e^1=\frac{1}{\mu}d\theta,\quad e^2=\frac{1}{\mu}\sin\theta d\varphi.
\end{align}
Then the spin connection is calculated as
\begin{align}
&\omega_{12}=-\cos\theta d\varphi, \quad \text{others}=0.
\end{align}
We take $SO(1,2)$ gamma matrices, which satisfy $\{\gamma^a,\gamma^b\}=2\eta^{ab}$, as
\begin{align}
\gamma^0=i\sigma_y,\quad \gamma^1=\sigma_x,\quad \gamma^2=\sigma_z,
\end{align}
where $\sigma_{x,y,z}$ are Pauli matrices.
Note that
\begin{align}
\gamma^a\gamma^b=\eta^{ab}+\epsilon^{ab}_{\phantom{ab}c}\gamma^c,
\end{align}
where $\epsilon^{abc}$ is the completely antisymmetric tensor satisfying $\epsilon^{012}=1$.
In this representation, spinors are real.
Let spinors and the gamma matrices have the following index structure:
$\psi_\alpha, \; (\gamma^a)_{\alpha}^{~\beta}$. 
We raise and lower the indices by the antisymmetric tensor 
$\epsilon^{\alpha\beta}$ and $\epsilon_{\alpha\beta}$
satisfying $\epsilon^{12}=-\epsilon_{12}=1$ as
$\psi^{\alpha}\equiv \epsilon^{\alpha\beta}\psi_\beta$ 
($\psi_{\alpha}= \epsilon_{\alpha\beta}\psi^\beta$),
$(\gamma^a)_{\alpha\beta}
\equiv \epsilon_{\beta\beta'}(\gamma^a)_\alpha^{\phantom{\alpha}\beta'}$ 
and
$(\gamma^a)^{\alpha\beta}
\equiv \epsilon^{\alpha\alpha'}(\gamma^a)_{\alpha'}^{\phantom{\alpha'}\beta}$.
The gamma matrices with two upper indices and two lower indices are symmetric:
$(\gamma^a)^{\alpha\beta}=(\gamma^a)^{\beta\alpha}$ 
and $(\gamma^a)_{\alpha\beta}=(\gamma^a)_{\beta\alpha}$.
We abbreviate the spinor indices for the following contractions:
\begin{align}
&\psi\chi\equiv \psi^\alpha \chi_\alpha=\chi\psi, \n 
&\psi\gamma^{a_1}\cdots\gamma^{a_k}\chi
\equiv \psi^\alpha(\gamma^{a_1}\cdots\gamma^{a_k})_{\alpha}^{\phantom{\alpha} \beta}\chi_\beta
\end{align}

%
%
%

\section{BPS solutions}
\setcounter{equation}{0}

In this appendix, we summarize the BPS solutions of $U(1)\times U(1)$ ABJM theory ($k>2$)
with respect to the cases in which $\eta_{AB}^{(+)}$ take  
\begin{align}
\text{(i)}:& \qquad 
\eta_{14}^{(+)}\neq 0 \;\; \text{and} \;\; \text{others}=0,  \n
\text{(ii)}:& \qquad
\eta_{14}^{(+)},\eta_{24}^{(+)} \neq 0 \;\; \text{and} \;\; \text{others}=0,  \n
\text{(iii)}:& \qquad
\eta_{14}^{(+)},\eta_{24}^{(+)},\eta_{34}^{(+)} \neq 0 \;\; \text{and} \;\; \text{others}=0.
\label{conditions for constant spinors}
\end{align}
Note that $\eta_{AB}^{(-)}=-\frac{1}{2}\epsilon_{ABCD}(\eta_{CD}^{(+)})^*$.
The other cases are essentially the same with one of these cases
(for instance, the case in which 
$\eta_{12}^{(+)}\neq 0$ and others$=0$ is equivalent to the case (i).).
For nonzero constant spinors, 
we can further impose the following projection
\begin{align}
i\gamma^0 \eta_{A'4}^{(+)}=s_{A'}\eta_{A'4}^{(+)},
\label{projection for eta+}
\end{align}
where $s_{A'}=\pm 1$. 
The projection for $\eta_{AB}^{(-)}$ is given by
\begin{align}
i\gamma^0 \eta_{A'B'}^{(-)}=s'_{A'B'}\eta_{A'B'}^{(-)},
\label{projection for eta-}
\end{align}
with $s'_{12}=s'_{21}=-s_{3},s'_{13}=s'_{31}=-s_{2},s'_{23}=s'_{32}=-s_{1}$.
The number of supersymmetries preserved 
for each case in \eqref{conditions for constant spinors} with and without 
\eqref{projection for eta+} and \eqref{projection for eta-} is summarized in Table 1.
From \eqref{del psi U(1)} one can easily get the BPS configurations 
of scalar fields for each case and then those of gauge fields from \eqref{EOM}.
Below we show the BPS solutions of scalar fields for each case.

\begin{table}
\begin{center}
\begin{tabular}{|c||c|c|}
\hline 
   & without \eqref{projection for eta+} and \eqref{projection for eta-} 
& with \eqref{projection for eta+} and \eqref{projection for eta-} \\ \hline
(i) & 4 & 2 \\ \hline
(ii) & 8 & 4 \\ \hline
(iii) & 12 & 6 \\ \hline
\end{tabular}
\caption{The number of supersymmetries for each BPS condition in ABJM on 
$\bR\times \bS^2$ ($k>2$): 
(i) $\eta^{(+)}_{14}\neq 0$ and $\eta^{(+)}_{24}=\eta^{(+)}_{34}=0$, 
(ii) $\eta^{(+)}_{14}, \eta^{(+)}_{24}\neq 0$ and $\eta^{(+)}_{34}=0$, and
(iii) $\eta^{(+)}_{14}, \eta^{(+)}_{24}, \eta^{(+)}_{34}\neq 0$.}
\end{center}
\label{BPS}
\end{table}

In the case (i) with \eqref{projection for eta+} and \eqref{projection for eta-},
\eqref{del psi U(1)} reduces to the following equations:
\begin{align}
&\partial_tY^{\overline{A}}+i\frac{\mu}{2}Y^{\overline{A}}
+s_1\mu \partial_\varphi Y^{\overline{A}}=0, \n
&\partial_\theta Y^{\overline{A}}+is_1\cot\theta \partial_\varphi Y^{\overline{A}}=0, \n
&\partial_tY^{\underline{A}}-i\frac{\mu}{2}Y^{\underline{A}}
+s_1\mu \partial_\varphi Y^{\underline{A}}=0, \n
&\partial_\theta Y^{\underline{A}}-is_1\cot\theta \partial_\varphi Y^{\underline{A}}=0,
\label{case (i) BPS equations}
\end{align}
where $\overline{A}=1,4$ and $\underline{A}=2,3$. These are easily solved as
\begin{align}
Y^{\overline{A}}
&=\sum_{p\in \bm{Z}_{\geq 0}}
v^{\overline{A}}_p \sin^p\theta e^{ip(s_1\varphi-t)-i\frac{\mu t}{2}}, \n
Y^{\underline{A}}
&=\sum_{p\in \bm{Z}_{\geq 0}}
v^{\underline{A}}_p \sin^p\theta e^{-ip(s_1\varphi-t)+i\frac{\mu t}{2}},
\label{case (i) BPS solutions}
\end{align}
where $v^{\overline{A}}_p$ and $v^{\underline{A}}_p$ are arbitrary constants. 
Note that if $Y^{\underline{A}}=0$ ($v^{\underline{A}}_p=0$) 
then $p$ of $v^{\overline{A}}_p$ can take values in $\bm{Z}_{\geq 0}+\frac{n}{k}$, 
where $n$ is an integer with $0\leq n<k$, because of the identification \eqref{orbifold}:
\begin{align}
Y^{\overline{A}}
&=\sum_{p\in \bm{Z}_{\geq 0}+\frac{n}{k}}
v^{\overline{A}}_p \sin^p\theta e^{ip(s_1\varphi-t)-i\frac{\mu t}{2}}, \n
Y^{\underline{A}}&=0.
\label{case (i) BPS solutions Y^{2,3}=0}
\end{align}
Without \eqref{projection for eta+} and \eqref{projection for eta-},
the BPS equation becomes \eqref{case (i) BPS equations} 
with the coefficient of $s_1$ being zero, so that
the corresponding BPS solution is $p=0$ solution in \eqref{case (i) BPS solutions}.

In the case (ii) with \eqref{projection for eta+} and \eqref{projection for eta-},
the BPS solution is given, only when $s_1=s_2$, by
\begin{align}
Y^1&=Y^2=0, \n
Y^{4}
&=\sum_{p\in \bm{Z}_{\geq 0}}
       v^{4}_p \sin^p\theta e^{ip(s_1\varphi-t)-i\frac{\mu t}{2}}, \n
Y^{3}
&=\sum_{p\in \bm{Z}_{\geq 0}}
       v^{3}_p \sin^p\theta e^{-ip(s_1\varphi-t)+i\frac{\mu t}{2}}.
\label{case (ii) BPS solutions}
\end{align}
The BPS solution without \eqref{projection for eta+} and \eqref{projection for eta-}
is the solution with $p=0$ in \eqref{case (ii) BPS solutions}.

In the case (iii) with \eqref{projection for eta+} and \eqref{projection for eta-},
the BPS solution is given, only when $s_1=s_2=s_3$, by
\begin{align}
Y^1&=Y^2=Y^3=0, \n
Y^{4}
&=\sum_{p\in \bm{Z}_{\geq 0}+\frac{n}{k}}
       v^{4}_p \sin^p\theta e^{ip(s_1\varphi-t)-i\frac{\mu t}{2}},
\label{case (iii) BPS solutions}
\end{align}
where we have taken into account the identification \eqref{orbifold}, 
so that $p$ can take an integer of $\bm{Z}_{\geq 0}+\frac{n}{k}$. 
The BPS solution without \eqref{projection for eta+} and \eqref{projection for eta-}
is the solution with $p=0$ in \eqref{case (iii) BPS solutions}.

\section{$\cN=8$ SYM on $\bR\times \bS^2$}
\setcounter{equation}{0}

In this appendix, we summarize $\cN=8$ SYM on $\bR\times \bS^2$.
The action of $\cN=8$ SYM on $\bR\times \bS^2$ is given by
\begin{align}
S_{SYM}&=\frac{1}{g_{SYM}^2}\int dt\frac{d\Omega}{\mu^2} \Tr\Biggl(
-\frac{1}{4}F^{ab}F_{ab}-\frac{1}{2}D'_{a}\phi {D'}^{a}\phi-\frac{\mu^2}{2}\phi^2
+\mu \phi F_{12} \n
& \qquad
-\frac{1}{2}D'_{a}X_{AB}{D'}^{a}X^{AB}-\frac{\mu^2}{8}X_{AB}X^{AB}
+\frac{1}{2}[\phi,X_{AB}][\phi,X^{AB}] 
+\frac{1}{4}[X_{AB},X_{CD}][X^{AB},X^{CD}]
\n
& \qquad
+i\hat{\psi}_A^\dagger \gamma^a D'_a \hat{\psi}^A 
+\frac{\mu}{4}\hat{\psi}_A^\dagger\gamma^0\hat{\psi}^A
\n 
& \qquad
-i\hat{\psi}_A^\dagger[\phi,\hat{\psi}^A]
-i\hat{\psi}_A^\dagger[X^{AB},\hat{\psi}_B^\dagger]
+i\hat{\psi}^{A}[X_{AB},\hat{\psi}^B]
\Biggr),
\end{align}
where $D'_a=\nabla_a+i[A_a,\cdot]$.
This theory is invariant under the following supersymmetry transformation
\begin{align}
\delta A^a
&=i\varepsilon_A^\dagger\gamma^{a}\hat{\psi}^A+
i\varepsilon^{A}\gamma^{a}\hat{\psi}_A^{\dagger},\n
\delta \phi
&=\varepsilon_A^\dagger\hat{\psi}^A-\varepsilon^{A}\hat{\psi}_A^{\dagger}, \n
\delta X^{AB}
&=\epsilon^{ABCD}\varepsilon_C^\dagger\hat{\psi}_D
-\varepsilon^{A}\hat{\psi}^B+\varepsilon^{B}\hat{\psi}^A,\n
\delta \hat{\psi}^A
&=-iD'_a\phi\gamma^a\varepsilon^A+\sum_{i=1,2}F_{0i}\gamma^{0i}\varepsilon^A
-2iD'_aX^{AB}\gamma^a\epsilon^*_B \n
&\quad
+(F_{12}-\mu\phi)\gamma^{12}\varepsilon^A
+\mu X^{AB}\gamma^{12}\varepsilon^*_B
+2i[\phi,X^{AB}]\varepsilon_B^*+2i[X^{AB},X_{BC}]\varepsilon^C
\label{SUSY transformation of SYM}
\end{align}
Here $\varepsilon^A$ are supersymmetry parameters which are 
$(1+2)$-dimensional Majorana spinors in the fundamental representation ($\bm{4}$) 
of $SU(4)$ given by
\begin{align}
\varepsilon^A
=e^{i\frac{\mu t}{4}}
e^{-i\frac{\theta}{2}\gamma^2}e^{\frac{\varphi}{2}\gamma^0}\varepsilon_0^A,
\end{align}
where $\varepsilon_0^A$ is a constant spinor.
$\varepsilon^*_A$ are the complex conjugate of $\varepsilon^A$
 and transform as the anti-fundamental representation of $SU(4)$.

The vacuum configuration of this theory is determined by the following equations
\begin{align}
&F_{12}-\mu\phi=0,\n
&D'_1\phi=D'_2\phi=0.
\end{align}
In the gauge in which $\phi$ is diagonal and $A_1=0$, these equations are solved 
by introducing two patches on $\bS^2$ as
\begin{align}
\phi&=\mu \: \diag\left(q_1,q_2,\cdots, q_N\right), \n
A_1&=0, \n
A_2&=\frac{1\pm \cos\theta}{\sin\theta} \phi,
\end{align}
where the upper and lower signs in $A_2$ 
correspond to the region I $(0\leq \theta <\pi)$ 
and the region I\hspace{-.1em}I $(0<\theta\leq \pi)$, respectively. 
The gauge field configuration for each diagonal component 
is Dirac monopole with magnetic charge $q_i$.
In the overlapping region of the region I and the region I\hspace{-.1em}I, 
the configurations on each patch are transformed each other by the transition function
\begin{align}
V_{\mathrm{I}\rightarrow \mathrm{I\hspace{-.1em}I}}
=\exp\left(i\frac{2}{\mu}\phi \cdot \varphi\right)
\end{align}
The single-valuedness of the transition function requires $q_i$ to be half-integer: 
$q_i\in \bZ/2$.

\section{Relation of fermions in ABJM and SYM}
\setcounter{equation}{0}

In this appendix, we explain in detail the interchange of $\psi_4$ and $\psi^{\dagger 4}$ 
\eqref{interchange} in the ABJM theory, 
which is needed for matching the ABJM theory (after the Higgsing) to $\cN=8$ SYM.
It is worthwhile to understand this interchange 
in terms of Clifford algebra representations of $SO(6)$ and $SO(8)$.
Let $\bar{\Gamma}^{I'}\: (I'=1,2,\cdots,6)$ be gamma matrices of $SO(6)$ satisfying
$\{\bar{\Gamma}^{I'},\bar{\Gamma}^{J'}\}=2\delta^{I'J'}$
and $\alpha^{A'} = \frac{1}{2}(\bar{\Gamma}^{A'} +i\bar{\Gamma}^{A'+3})$ 
and $\alpha_{A'}^\dagger = \frac{1}{2}(\bar{\Gamma}^{A'} -i\bar{\Gamma}^{A'+3})$.
$\alpha^{A'}$ and $\alpha_{A'}^\dagger$ satisfy 
$\{\alpha^{A'},\alpha_{B'}^\dagger\}=\delta^{A'}_{B'}$ 
and are regarded as annihilation and creation operators of fermions 
on the Fock vacuum $|\bar{\Omega}\rangle$.
Note that the $U(3)$ rotation defined by 
$\alpha^{A'} \rightarrow (U^*)^{A'}_{~B'}\alpha^{B'}$
and $\alpha_{A'}^\dagger \rightarrow U_{A'}^{~B'}\alpha_{B'}^\dagger$
 is a subgroup of $SO(6)$.
The (Dirac) spinor representation of $SO(6)$ is expressed as
\begin{align}
\bm{8}=\{|\bar{\Omega}\rangle,\: 
\alpha_{A'}^\dagger|\bar{\Omega}\rangle, \:
\alpha_{A'}^\dagger \alpha_{B'}^\dagger|\bar{\Omega}\rangle, \:
\alpha_{A'}^\dagger \alpha_{B'}^\dagger \alpha_{C'}^\dagger|\bar{\Omega}\rangle\},
\end{align}
One can decompose $\bm{8}$ in terms of the eigenvalue of the chirality matrix 
$\bar{\Gamma}=\prod_{I'=1}^{6}\bar{\Gamma}^{I'}=\prod_{A=1}^{4}(1-2\alpha_A^\dagger \alpha^A)$
into two Weyl representations as
\begin{align}
\bm{8}\rightarrow \bm{4}+\bar{\bm{4}}
\end{align}
where
\begin{align}
\bm{4}&=\left\{\alpha_{A'}^\dagger|\bar{\Omega}\rangle,\: 
\alpha_{A'}^\dagger \alpha_{B'}^\dagger \alpha_{C'}^\dagger|\bar{\Omega} \right\}, \n
\bar{\bm{4}}&=\left\{|\bar{\Omega}\rangle,\:
\alpha_{A'}^\dagger \alpha_{B'}^\dagger|\bar{\Omega}\rangle \right\}.
\end{align}
and $\bm{4}$ and $\bar{\bm{4}}$ have $\bar{\Gamma}=1$ and $\bar{\Gamma}=-1$, respectively.
We further decompose $\bm{4}$ and $\bar{\bm{4}}$ of $SU(4)$ into $SU(3)\times U(1)$ 
where the $U(1)$ charge is specified by $\sum_{A'=1}^{3}[\alpha^{A'},\alpha_{A'}^\dagger]/2$:
\begin{align}
\bm{4}&\rightarrow \bm{3}_{1/2} + \bm{1}_{-3/2}, \n
\bar{\bm{4}}&\rightarrow \bar{\bm{3}}_{-1/2} + \bm{1}_{3/2}.
\label{SU(4) into SU(3)xU(1) in SYM}
\end{align}

Next, let $\Gamma^I \: (I=1,2,\cdots, 8)$ be the gamma matrices of $SO(8)$ satisfying
$\{\Gamma^I,\Gamma^J\}=2\delta^{IJ}$
and $\beta^{A} = \frac{1}{2}(\Gamma^{A} +i\Gamma^{A+4})$ 
and $\beta_{A}^\dagger = \frac{1}{2}(\Gamma^{A} -i\Gamma^{A+4})$.
$\beta^A$ and $\beta_A^\dagger$ satisfy $\{\beta^A,\beta_B^\dagger\}=\delta^A_B$ 
and are regarded as annihilation and creation operators of fermions 
on Fock vacuum $|\Omega\rangle$.
By using the fermion Fock space,
the (Dirac) spinor representation of $SO(8)$, $\bm{16}$, is given as: 
\begin{align}
\bm{16}=\{|\Omega\rangle, 
\beta_{A}^\dagger|\Omega\rangle, \:
\beta_{A}^\dagger \beta_{B}^\dagger|\Omega\rangle, \:
\beta_A^\dagger \beta_B^\dagger \beta_C^\dagger|\Omega\rangle, \: 
\beta_A^\dagger \beta_B^\dagger \beta_C^\dagger \beta_D^\dagger|\Omega\rangle\}.
\end{align}
In terms of the eigenvalue of the chirality matrix 
$\Gamma\equiv \prod_{I=1}^{8}\Gamma^I=\prod_{A=1}^{4}(1-2\beta_A^\dagger \beta^A)$,
$\bm{16}$ is decomposed as 
\begin{align}
\bm{16}\rightarrow \bm{8}_s + \bm{8}_c, 
\end{align}
where
\begin{align}
\bm{8}_s
&=\left\{
\beta_{A}^\dagger|\Omega\rangle,\: 
\beta_A^\dagger \beta_B^\dagger \beta_C^\dagger|\Omega\rangle
\right\}, \n
\bm{8}_c
&=\left\{
|\Omega\rangle,\: \beta_{A}^\dagger \beta_{B}^\dagger|\Omega\rangle,\: 
\beta_A^\dagger \beta_B^\dagger \beta_C^\dagger \beta_D^\dagger|\Omega\rangle
\right\},
\end{align}
and $\Gamma=-1$ for $\bm{8}_s$ and $\Gamma=1$ for $\bm{8}_c$.
We decompose these into $SU(4)\times U(1)$ where the $U(1)$ charge specified by
$\sum_{A=1}^{4}[\beta^A,\beta_A^\dagger]/2$.
In particular, $\bm{8}_s$ is decomposed as
\begin{align}
\bm{8}_s
&\rightarrow 
\bm{4}'_{1}+\bar{\bm{4}}'_{-1},
\end{align}
where
\begin{align}
\bm{4}'_{1}&=\left\{\beta_{A}^\dagger|\Omega\rangle\right\}, \n
\bar{\bm{4}}'_{-1}
&=\left\{\beta_A^\dagger \beta_B^\dagger \beta_C^\dagger|\Omega\rangle\right\}.
\end{align}
We further decompose $SU(4)$ into $SU(3)\times U(1)$ as before
with the $U(1)$ charge specified by $\sum_{A'=1}^{3}[\beta^{A'},\beta_{A'}^\dagger]/2$:
\begin{align}
\bm{4}'&\rightarrow \bm{3}_{1/2}+\bm{1}_{3/2}, \n
\bar{\bm{4}}'&\rightarrow \bar{\bm{3}}_{-1/2}+\bm{1}_{-3/2}.
\label{SU(4) into SU(3)xU(1) in ABJM}
\end{align}
We then see that the two sets, 
\eqref{SU(4) into SU(3)xU(1) in SYM} and \eqref{SU(4) into SU(3)xU(1) in ABJM} 
are not in one to one correspondence with each other. 
In particular to identify the fermions of the 
ABJM theory with the fermions of the SYM (after Higgsing), 
we must interchange  $\bm{1}_{3/2} \leftrightarrow \bm{1}_{-3/2}$. 
This corresponds to interchanging $\psi_{4}\leftrightarrow \psi^{4\dagger}$ in the ABJM.

\end{document}